\documentclass[prb,twocolumn,showpacs,preprintnumbers,amsmath,amssymb]{revtex4}
\usepackage{bm}
\usepackage{graphicx}
\usepackage{epstopdf}

 \def \S {$S/N $ }
  \def \Ga {$^{71}$Ga }
   
  \def \GAl{Al$_{0.13}$Ga$_{0.87}$As }
   \def \ie {{\it i.e.}, }
 \begin{document}

\title{NMR Study of Large Skyrmions in Al$_{0.13}$Ga$_{0.87}$As Quantum Wells}
\author{V. F. Mitrovi{\'c}$^{1,2}$, M. Horvati{\'c}$^{1}$,  C. Berthier$^{1}$, S. A. Lyon$^{3}$, M. Shayegan$^{3}$}
\address{$^{1}$ Grenoble High Magnetic Field Laboratory,
MPI-FKF and CNRS, B.P. 166, 38042 Grenoble
Cedex 9, France\\
$^{2}$  Department of Physics, Brown University,
Providence, RI 02912, U.S.A. \\
$^{3}$Department of Electrical Engineering, Princeton University,
Princeton, NJ 08544, U.S.A.}

\date{\today}

\begin{abstract}
A nuclear magnetic resonance (NMR) study  is reported of   multiple (30) 
Al$_{0.13}$Ga$_{0.87}$As quantum well  (QW) sample near the Landau level filling factor 
$\nu =1$. In these \GAl QWs  the effective $g$ factor is nearly zero. This can lead to two effects: vanishing electronic
polarization $(\mathcal{P})$ and skyrmionic excitations composed of a huge
number of spins. As small $\mathcal{P}$ values cause an overlap of the NMR signals
from the QW and barriers, a  special technique was employed to allow 
 these two signals to be distinguished. The
QW   signal corresponds to a small, negative, and very broad distribution of
spin polarization  that exhibits thermally induced depolarization. Such a distribution can be
attributed to sample inhomogeneities and/or to large skyrmions,  the latter
possibility being favored by observation of a very fast $T_{2}^{-1}$ rate. 
 
\end{abstract}

 \maketitle



\section{Introduction}

In two-dimensional electron systems (2DESs), the ground state at the Landau
level filling factor $\nu =1$ is a ferromagnetic state where only the
lowest Landau level is completely filled with spin-up electrons \cite{Gervin}. This spin-polarized state of a quantum Hall ferromagnet is particularly
interesting because the low-lying charged excitations are skyrmions, complex
charged spin texture \cite{Sondhi93, Bray95,Cooper97,Lejnell}. The spin texture of a
skyrmion encompasses many progressively reversed spins. Its size, $S$, is
defined as the number of reversed spins within an elementary excitation. It
is governed by the ratio, $\eta \equiv E_{Z}/E_{C}$, of the Zeeman energy, 
$E_{Z}=g\mu _{B}H$, limiting the number of spin flips, and the
Coulomb interaction energy, favoring local ferromagnetic ordering
\cite{Lejnell}. Therefore, reducing $E_{Z}$ towards zero leads to
divergence of skyrmion size. One way to achieve this interesting
limiting regime is to lower the effective $g$ factor to nearly
zero. This can be achieved either by the application of
hydrostatic pressure, or as in the case of our sample by confining 
the 2DES to an Al$_{x}$Ga$_{1-x}$As  quantum well (QW) where the Al composition
 $(x)$  is $\simeq 0.13$.

In pure GaAs QWs, where $g\approx -0.44$, a skyrmion size in the range 
$3.6<S<9$ was inferred using different experimental techniques 
\cite{Barrett95, Schmeller95, Aifer96, Bayot96, Bayot97, Melinte99}. A large skyrmion size of $S=36$
was deduced from   magnetotransport measurements under pressure
 \cite{Nicholas95, Maude96, Leadley98}, in which the limit of
$g \rightarrow 0$ was reached. Although investigation  of $S$ using
pressure seems to be very convenient for systematic exploration
of the $g$ dependence, it is compromised by two requirements: for
each $g$ (pressure) value, a separate cooldown is necessary and this leads to different
disorder, and illumination of the sample is necessary to
compensate for the loss of the density of 2DES induced by the
application of pressure. An alternative possibility is to tune $g$
to zero by confining the 2DES to \GAl QWs  \cite{Weisbuch77, Shukla00}.
Using this approach, from their magnetotransport measurements in
Al$_{0.13}$Ga$_{0.87}$As QW, Shukla \textit{et al.}
\cite{Shukla00} reported the largest ever skyrmions size $S\approx
50$.

Nuclear magnetic resonance (NMR) has proved to be a powerful tool for confirming the existence of
skyrmionic excitations \cite{Barrett95}. In addition, it has provided
valuable information about the microscopic nature and dynamics of
this many body electronic state \cite{Barrett95, Melinte,
Khandelwal01}. Therefore, the NMR technique appears to be a good
candidate for shedding light on the nature of the large skyrmions in \GAl QWs.  Here, we present   such an investigation.

We studied the $^{71}$Ga NMR signal in a multiple
Al$_{0.13}$Ga$_{0.87}$As QW sample, of the same composition as the
one studied in Ref.\cite{Shukla00},  at  $\nu =1$. The electronic polarization $(\mathcal{P})$ is
found to be very small, causing the overlap of signals from the QWs 
and barriers. The small $\mathcal{P}$ would also inhibit the enhancement of the weak
signal from QWs by optically pumped NMR, justifying our use of
``conventional NMR''. A special technique was employed to  
distinguish between a weak signal from the QWs and one from the
barriers. In these QWs  we observe the thermally induced depolarization of
  small and negative (compared to a pure GaAs QW)
  spin polarizations. We argue that this can be
attributed to large skyrmions. 

The paper is organized as follows. In \mbox{Sec. \ref{Exptech}} experimental
details are presented. These include descriptions of the sample,
experimental setup, and essentials of NMR in QWs. The small tip angle
technique is described in \mbox{Sec. \ref{TipAngle}}. Our findings are
summarized in \mbox{Sec. \ref{ResultsS}}. Finally, implications of these on the
nature of elementary excitations are discussed in \mbox{Sec. \ref{Disc}}.

\section{Experiment}

\label{Exptech}

\subsection{Experimental Technique and the Sample}

We   investigated a multiple-QW sample, similar to the single QW sample used in
transport measurements \cite{Shukla00}. The results  indicate that
the sample is of   high quality and  that its $g$ factor is small
\cite{Shukla00}. Our sample consists of thirty 24 nm-wide Al$_{0.13}$Ga$_{0.87}$As QWs
bounded on each side by 132 nm-thick Al$_{0.35}$Ga$_{0.65}$As `barriers' which are
Si doped near their centers.  The density of the 2D electrons
confined to each of the QWs is \mbox{ $n_{2D}  = 1.1 \times 10 ^{11}\,
\rm{cm}^{-2}$} and their mobility is 
\mbox{ $\mu_0  = 3 \times 10 ^4 \,
\rm{cm}^2 \rm{/Vs}$}. The low mobility is believed to be predominantly due to impurity
rather than alloy scattering \cite{Shukla00}. The effective $g$ factor  for the 2D electrons in our
   sample is estimated to be $g\approx +0.04\,$ \cite{Ivchenko92,
Kiselev98, Shukla00}. It should be noted that this  is of the opposite sign and an order of magnitude
smaller than that in the bulk GaAs.

The NMR spectra were recorded using a custom built NMR
spectrometer. For temperatures $(T)$ above 1 K we were not able to
separate the QW signal due to insufficient signal-to-noise ($S/N$)  ratio in this sample.
The low temperature environment, down to 50 mK, was provided by a
$^3$He/$^4$He dilution refrigerator. The RF coil was mounted into
the mixing chamber of the refrigerator. This ensured good thermal
contact with the sample. We   used the `bottom-tuning' scheme in
which a variable tuning capacitor was mounted as close as
possible to the coil just outside of the mixing chamber. This
tuning scheme minimizes the RF-losses, and so optimizes the NMR
sensitivity. It allows a clear separation of the QW spectra at
all temperatures below 1 K, otherwise impossible for \mbox{$ T
\gtrsim 200$ mK}.

\subsection{NMR in Quantum Wells}

\label{NMRQW}

In pure GaAs QW samples,  previous optically pumped \cite{Barrett95} and
conventional NMR \cite{Melinte} studies have shown that NMR spectra consist of
two well separated peaks originating from the nuclei located in the QWs and the
barriers. These peaks are distinguishable owing to a significant
polarization of the electrons in the QWs. This is because the hyperfine
interaction between nuclei and electrons gives rise to the Knight shift, $K_{S}$,
of the NMR line \cite{AbragamBook, Richardson86, Melinte}
\begin{equation}
K_{S}({\mathbf{R}}_{i})\propto \mathcal{P}({\mathbf{R}}_{i})\rho _{e}({%
\mathbf{R}}_{i}).
\end{equation}%
Here $\mathcal{P}(R_{i})$ is the local spin polarization at the spatial
position, ${\mathbf{R}}_{i}$, of the nuclei, and $\rho _{e}$ the local value of
the electron density. Signal from   nuclei in the barriers, where the
electron density is vanishingly small, will be essentially unshifted, and
can be used as a zero-shift reference. The nuclei in the QWs provide an NMR line
whose shift is directly proportional to the average global electron
polarization,
\begin{equation}
K_{S}(T)=A_{c}\rho _{e}(0)\mathcal{P}(T),  \label{ShPol2}
\end{equation}%
assuming uniform electronic density, $\rho _{e}(0)$. The effective hyperfine
coupling constant, $A_{c}$, can be determined experimentally from the NMR
shift obtained at low temperature, high field, and at filling factors, such
as $\nu =1/3$, where the 2DES is fully polarized, i.e., $\mathcal{P}=1$ and $%
K_{S}$ reaches its maximal value. From the values observed in pure
GaAs multiple QW samples\cite{Barrett95, Melinte}, we infer that
a maximum shift of $K_{S}^{\mathcal{P}=1}  \simeq$  24 kHz is expected for our sample.
This allows for direct deduction of the spin polarization of 2DES
from $\mathcal{P}=K_{S}(T)/K_{S}^{\mathcal{P}=1}$.
\begin{figure}[t]
\centerline{\includegraphics[scale=0.47]{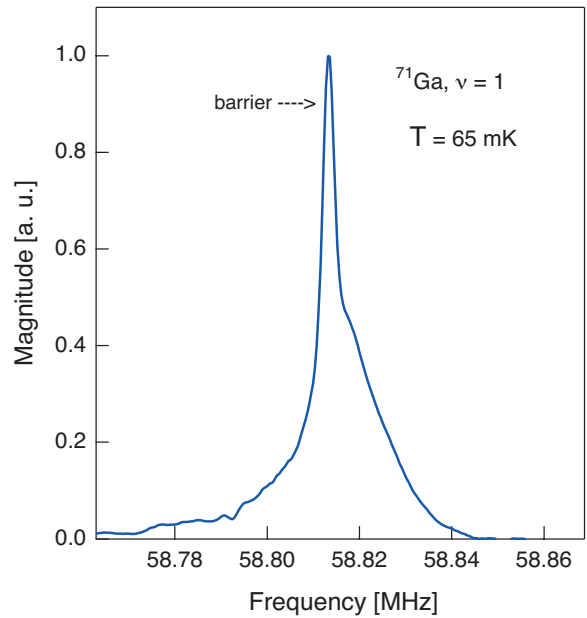}}
\begin{minipage}{0.93\hsize}
\caption[]{\label{RowData}\small (Color online) The \Ga NMR spectrum, containing
the signal from  nuclei in both QWs and barriers, measured at $\nu
= 1$ and $T = 65$ mK. The spectrum was recorded using the
effective pulse length and the repetition times that favor the
barriers' signal (sharp narrow peak).}
\end{minipage}
\end{figure}
\noindent

However, if the effective electronic $g$ factor is lowered,  the absolute
polarization, $|\mathcal{P}(g\mu_B H /k_B T)|$,  is smaller for the same value of $H$ and $T$.  As a consequence, the  separation between the barrier  and
QW signals in NMR spectra is  smaller as well. 
Indeed,  as illustrated in \mbox{Fig. \ref{RowData}}, the NMR spectrum of the sample exhibits an 
asymmetric peak which contains contributions from the \GAl QWs (a broad, background-like signal) 
and from the barriers (a strong, narrow peak). 
 Nevertheless, the two contributions can be discerned by exploiting their
different nuclear spin-lattice relaxation times ($T_{1}$). The hyperfine
interaction responsible for the shift of the NMR line induces the $T_{1}$
relaxation as well \cite{AbragamBook}. It contains  spin operators of the
form $I_{\pm }S_{\pm }^{e}({\mathbf{R}}_{i})$ which produce simultaneous
electronic, $S^{e}$, and nuclear spin, $I$, flips. In addition, the
existence of closely spaced electron levels with energy separation
comparable to the nuclear Zeeman gap is required by  energy conservation.
Such levels are provided by the additional degrees of freedom present in a real
2DES. These degrees of freedom could be generally ascribed to disorder
and/or the presence of skyrmions in the electronic ground state in the
vicinity of $\nu =1$ \cite{Vagner88, Vagner95, Antoniou, Cote97,Green00}.
Therefore, even in the limit of   vanishing $g$ factor, we expect
significantly faster $T_{1}$ relaxation of the nuclei in the QW compared to
those in the barriers, due mostly to the existence of low energy excitations,
such as skyrmions. This is further  enhanced by the thermal
depolarization.

The NMR spectra are recorded by the small tip angle free induction
decay (FID) technique that we describe in detail in \mbox{Sec.
\ref{TipAngle}}. The essence of this technique is that the signal
intensity depends strongly on the effective pulse strength and the
ratio of the repetition time between consecutive spectrum
acquisitions  and the relaxation time. Thus, one can find the tip
angle that maximizes the $S/N$  ratio, referred
to as the \textit{Ernst angle} \cite{CowanBook, Ernst} (\mbox{Eq.
\ref{ThMax}} in the following sub-section). The two contributions
to the NMR spectrum, having different relaxation rates, have different
Ernst angles, as shown in \mbox{Fig. \ref{SN_Pi2TrT1}}.
Therefore, we can obtain the pure QW spectrum from the analysis of
the pulse strength dependence of the lineshape. The effective
pulse strength, i.e. total pulse power, was varied in two ways: by
varying the time of the pulse duration and by varying the voltage
of the pulse, keeping its time duration fixed. The latter method
is preferred since it eliminates the possible artifacts associated
with the variable bandwidth coverage of the pulse.

We remark here that the small $g$ factor also inhibits the
application of the optical pumping technique. Thus, the only way to
separate signal from nuclei in QWs and so
probe the $\mathcal{P}$ in these samples is by the conventional NMR,
exploiting the different $T_{1}$\ relaxation times of nuclei in the QWs
and   barriers. The technique
providing this ``$T_{1}$ contrast'' is described in detail in the
next sub-section.


\subsection{Small Tip Angle Pulse Technique}

\label{TipAngle}

The small tip angle pulse method exploits the fact that the intensity of the
NMR spectra depends strongly on the effective power $(i.e.,\, {\rm the \, tip\, angle}\, \theta )$ of
the excitation pulse and the ratio of the repetition time $(T_{R})$ between
consecutive spectrum acquisitions and the relaxation time $(T_{1})$. In a
conventional NMR experiment each data acquisition consists of an excitation
and detection of magnetization, and usually enough time is allowed between
acquisitions that the equilibrium spin temperature is fully restored, $%
T_{R}\geq 5T_{1}$. The spins are manipulated in several standard ways of
which the two simplest examples are: FID \mbox{$({\pi \over 2} -{\rm acquire})$}
and Hahn echo \mbox{$({\pi \over 2} - \tau - \pi-{\rm acquire})$} sequences.
Usually, these sequences are repeated many times with a delay of $T_{R}$ so that the signal is 
 averaged in order to improve the signal-to-noise ratio.

As the echo sequence is rather incompatible with the small tip angles, in
what follows we will only discuss the FID case   used in our experiment.
Following the exact $\theta ={\frac{\pi }{2}}$ pulse the maximum signal
intensity is obtained if $T_{R}\geq 5T_{1}$. In general, if the pulse angle
is smaller and/or if one does not wait this long between consecutive
acquisitions the signal will be smaller. However, one can show that the $S/N$
ratio can be improved if short $T_{R}$ are used concomitantly with a smaller
tip angle, \ie if $\theta \lesssim {\frac{\pi }{2}}$ \cite{Ernst}. In this
case the benefit in \S from increasing the rate of data accumulation more
than makes up for loss of signal from not allowing the magnetization to
return to its thermal equilibrium value. Furthermore, the amplitude of the
signal depends strongly on the spin-lattice relaxation rate. Therefore,
tuning the excitation angle strength allows for separation of spectral
components with  distinct relaxation rates by optimizing respective
signal amplitudes, as illustrated in \mbox{Fig. \ref{VaryPi2_T65mK}}. This
is demonstrated in a more quantitative way in the next paragraph.

\begin{figure}[b]
\centerline{\includegraphics[scale=0.45,angle=0]{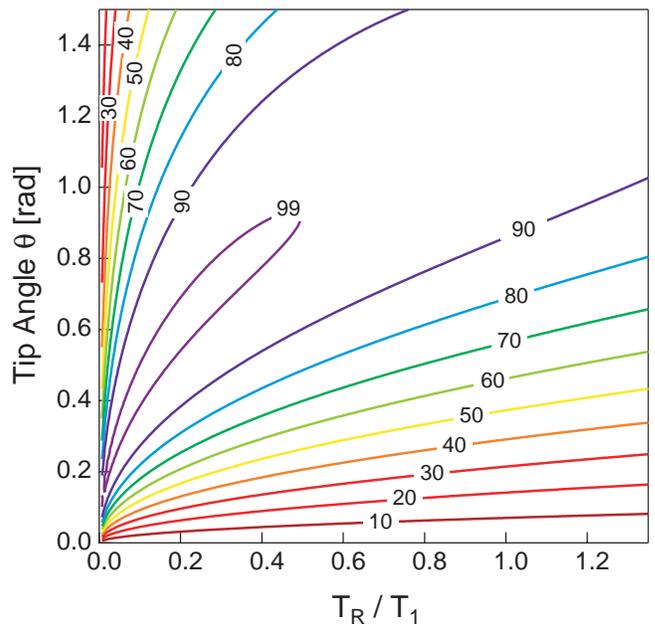}}
\begin{minipage}{0.93\hsize}
\caption[]{\label{SN_Pi2TrT1}\small (Color online) The contour-image plot  of the
signal-to-noise ratio as a function of an effective pulse tip
angle and  the  ratio of the repetition time, $T_R$, to the
spin-lattice relaxation time, $T_1$.  The relative value of the $S/N$ in percents is denoted on the contours.}
\end{minipage}
\end{figure}
\noindent

We consider the signal magnitude after consecutive data
acquisitions. Following an excitation pulse, the longitudinal
magnetization $M_{z}$ of a system of spin 1/2 nuclei recovers
towards its equilibrium value $M_{0}$ by exponential relaxation,
\begin{equation}
M_{z}(t)=M_{0}-\left( M_{0}-M_{z}(t=0)\right) e^{-t/T_{1}},
\end{equation}%
where $M_{z}(t=0)$ is the initial magnetization. After a time
$T_{R}$ the magnetization will recover to $M=M_{z}(T_{R})$, prior
to the next pulse which will rotate the magnetization by the angle
$\theta$ and restart the acquisition process. After this pulse the
longitudinal magnetization will be equal to $M \cos (\theta )$,
which is the starting magnetization of the following relaxation.
Thus, in the steady state the above equation (at $t=T_{R}$)
becomes \mbox{$ M=M_{0}-\left( M_{0}-M\cos (\theta )\right)
e^{-T_{R}/T_{1}}.$} Solving for $M$, one obtains
\begin{equation}
M=M_{0}\,{\frac{{1-e^{-T_{R}/T_{1}}}}{{1-\cos (\theta
)e^{-T_{R}/T_{1}}}}}\,.
\end{equation}%
In an experiment we detect a \textit{transverse} magnetization
just after the pulse, which is $M\sin (\theta )$. Thus, the signal
magnitude is proportional to
\[
S\propto {\frac{{1-e^{-T_{R}/T_{1}}}}{{1-\cos (\theta )e^{-T_{R}/T_{1}}}}}%
\,\sin (\theta ).
\]%
The noise is inversely proportional to the square root of the number
of averages $N$. Thus, in a given time it is proportional to
$\sqrt{T_{R}}$. The signal to noise ratio is thus
\begin{equation}
S/N\propto {\frac{{\left( 1-e^{-T_{R}/T_{1}}\right) }\sin (\theta )}{{%
\left( 1-\cos (\theta )e^{-T_{R}/T_{1}}\right) \sqrt{T_{R}}}}}\,.
\label{S/N}
\end{equation}%
In \mbox{Fig. \ref{SN_Pi2TrT1}} one can see that by concomitant
shortening of the tip angle and the repetition time, one can
maintain a high $S/N$\ ratio. From \mbox{Eq. \ref{S/N}}\ it is easy
to find that the   tip angle that maximizes the $S/N$ ratio is
given by
\begin{equation}
\cos (\theta )=e^{-T_{R}/T_{1}}\,.  \label{ThMax}
\end{equation}%
We remark that the detection of this angle allows one to determine  $T_{1} $ much 
 faster than in the standard measurement. However, a  $T_{1} $ value 
 determined in such a fashion certainly has higher error. In practice, in a non-homogeneous
and complicated system such as QW samples this method of $T_{1}$
determination turns out to be unsuitable.

\section{Results and Discussion}

\label{ResultsS}

\subsection{Signal Separation}

\label{SSep}
\begin{figure}[b]
\centerline{\includegraphics[scale=0.45]{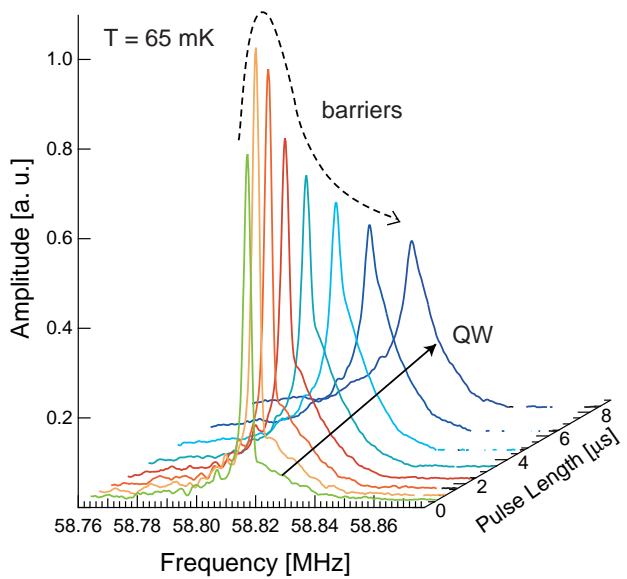}}
\begin{minipage}{0.93\hsize}
\caption[]{\label{VaryPi2_T65mK}\small (Color online)  The $^{71}$Ga NMR spectra  
 at  $\nu=1$ taken by   free induction decay. The effective pulse length
dependence of the spectra is used to distinguish the signal of
nuclei in  QWs from the signal of those in barriers. }
\end{minipage}
\end{figure}
\noindent

\begin{figure}[b]
\centerline{\includegraphics[scale=0.43]{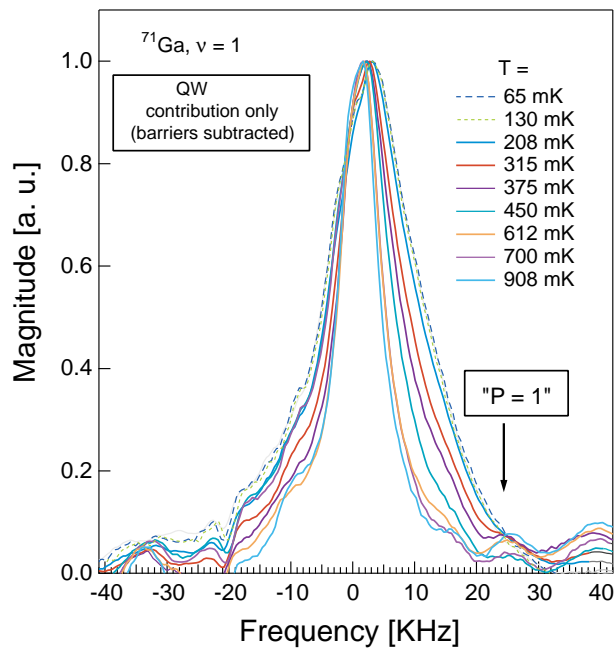}}
\begin{minipage}{0.93\hsize}
\caption[]{\label{GSubVaryT}\small (Color online) The temperature dependence of
the QW signal. The zero of the frequency (i.e., the spin
polarization) scale is defined by the position of the signal from
barriers containing no free electrons. The position of the spectra defines the Knight shift, $K_{S}(\nu=1)$. }
\end{minipage}
\end{figure}
\noindent

Full spectra at $\nu=1$ recorded using various effective pulse lengths at 65 mK are
shown in \mbox{Fig. \ref{VaryPi2_T65mK}}. It is evident that the spectral shape
is strongly dependent on the effective pulse angle, as previously discussed.
Nonetheless, two main features can be discerned: a narrow Gaussian-like
component at lower frequencies and a broad asymmetric component. The narrow
component dominates the spectra obtained using weak excitation pulses. This
implies that the signal comes from the nuclei with a very long relaxation
time, \ie$T_{1}\rightarrow \infty $, as suggested by \mbox{Eq. \ref{ThMax}}
and \mbox{Fig. \ref{SN_Pi2TrT1}}. Indeed, signal from the nuclei in the
barriers is expected to relax extremely slowly compared to the signal from
QW. This is due to lack of free electrons in the barriers. Furthermore, the
linewidth of this component is found to be  \mbox{2  kHz}, comparable to
well separated barrier signals  observed in other pure GaAs QW samples \cite{Barrett95}. On
the other hand, the broad asymmetric component dominates the spectra
obtained using stronger excitation pulses, implying that this signal comes
from the nuclei with   relatively short relaxation times. This is expected
for nuclei in QWs due to their hyperfine coupling with a significant number of
free electrons. Therefore, the signal from the QWs can be obtained using stronger
excitation pulses. This allows a separation of the two signals. The details of
the procedure will be discussed next.

We first record the spectra using weak excitation pulses to obtain the precise
spectral component from the barriers at a given temperature. The signal is
fitted to a Gaussian that is then subtracted from the spectra recorded using
strong excitation pulses. Spectra obtained in this manner contain only information from
 the  nuclei in QWs, as   illustrated in  \mbox{Fig. \ref{GSubVaryT}}.

\subsection{Electron Spin Polarization}

After subtraction of the barriers' contribution from the NMR spectra we are
left with a pure QW lineshape. Its temperature evolution is shown in 
 \mbox{Fig. \ref{GSubVaryT}}. We point out that the spectral position  defines the Knight shift, $K_{S}(\nu=1)$, as discussed in 
 \mbox{Sec. \ref{NMRQW}}.  
 However, the most striking observation is the appearance of the significant 
 linewidth broadening  with decreasing temperature. The broadening is asymmetric.
 The notable width of the lower temperature spectra implies a significant
distribution of the spin polarizations. Such a distribution can be attributed
to sample inhomogeneities and/or to large skyrmions. We will address this
issue in \mbox{Sec. \ref{Disc}}.

\begin{figure}[t]
\centerline{\includegraphics[scale=0.95]{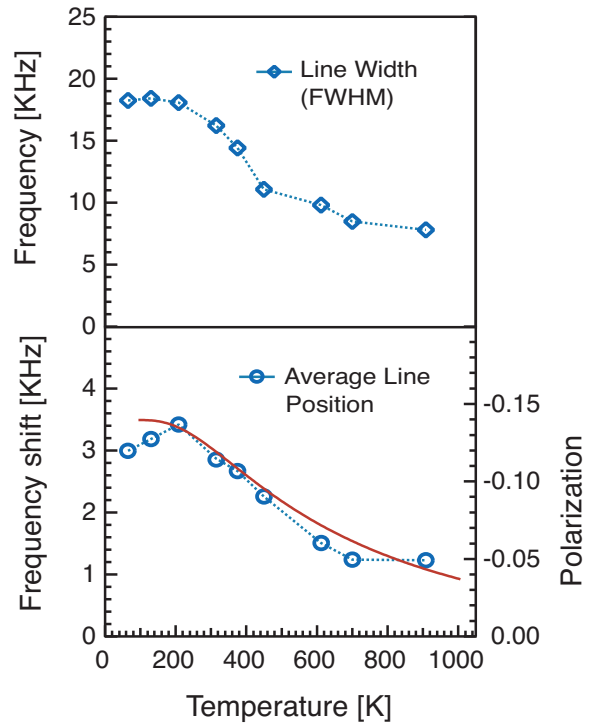}}
\begin{minipage}{0.95\hsize}
\caption[]{\label{GSubLW}\small (Color online)  The temperature dependence
(thermal depolarization) of the  linewidth and lineposition (the shift, $K_S$) of QW
signal at  $\nu=1$ shown in \mbox{Fig.\ref{GSubVaryT}}. The dotted lines are guide
to the eye. The solid line is the fit to $\tanh$ function as
described in the text.}
\end{minipage}
\end{figure}

In addition to the linewidth broadening, decreasing temperature
induces a shift of the average position of the QW signal. The
temperature dependence  of the linewidth and the lineposition are 
summarized in \mbox{Fig. \ref{GSubLW}}. We point out that the
frequency scales of the figures can be directly related to the
spin polarization, as the hyperfine coupling
constant is known from measurements in pure GaAs QW samples. For 
our sample \mbox{$K_{S}^{\mathcal{P}=1}=24$ kHz} corresponds to a full polarization,
as discussed in \mbox{Sec. \ref{NMRQW}}. It is evident from
\mbox{Fig. \ref{GSubLW}} that the polarization is very small,
$\approx 1/8$ of the full polarization, and \textit{negative}
compared to pure GaAs QWs.  This result is consistent with the small  value
of $g\approx +0.04$ reported in \mbox{Ref. \cite{Shukla00}} for a  sample of the same
Al concentration. One should note that this $g$ value is much smaller in magnitude compared to 
 $g\simeq -0.44$ in pure GaAs QWs and has the opposite sign. 
 
At the filling factor  in the vicinity of  but not at, $\nu =1$ the spin polarization is
expected to be significantly smaller than at $\nu =1$ in the presence of
large skyrmions \cite{Barrett95}. We did not observe such a drop in
polarization, possibly due to a small absolute value of the polarization at $%
\nu =1$. Therefore, direct deduction of the skyrmion size was
impossible. We remark that reduced spin polarization of the skyrmions 
   at $\nu =1$ has already been  reported even in pure GaAs
multiple QW samples \cite{Melinte} presenting otherwise full
polarization in the low-$T$ limit at $\nu =1/3,1/2$
and 2/3 \cite{Melinte, Freytag}.  The full polarization 
 $\mathcal{P}(T\rightarrow 0,\, \nu =1)=1$\ is approached only in very pure and homogeneous
samples, preferably with    smaller numbers of quantum wells\cite{Barrett95,
Khandelwal01}. The \GAl doped QWs are intrinsically less pure than pure GaAs
ones.  Thus, the observed reduction of $\mathcal{P}(T\rightarrow 0,\, \nu =1)$ is
indeed much stronger and essentially no significant $\nu-$dependence of the shift and of the linewidth is observed in the range of $\nu = 1 \pm 0.1$. 
For this reason,  a comparison of the $\nu$ dependence with the one observed in Ref. \cite{Khandelwal01} is inconclusive. 
\noindent

\subsection{Discussion}
\label{Disc}

Since the temperature dependence of the shift probes low-lying excitations, a
significant amount of theoretical work has been devoted to calculating the $T
$ dependence of the electron spin polarization precisely at $\nu =1$ \cite{Read95,Haussmann97, Green98,Rodriguez98, Chakraborty99, Kasner00, Brey00}.
In a simple model at $\nu =1$ for noninteracting 2D electrons with the
chemical potential in the middle of the Zeeman gap, $\mathcal{P}(T)$ is
given by
\begin{equation}
\mathcal{P}(T,\nu =1)=\tanh \left( \frac{E_{Z}}{4k_{B}T}\right) .
\end{equation} 
In \mbox{Fig. \ref{GSubLW}}, we plot a fit to the above equation with 
$E_{Z}$ as an adjustable parameter. Using the $E_{Z}$ value
obtained from the fit we find that $g=0.43\pm 0.10$. However, in
 pure GaAs QWs it was necessary to use a Zeeman splitting enhanced
by a factor of $\approx 10$ to fit the data \cite{Usher90,
Barrett95}. This enhancement reveals the importance of interaction
at \mbox{$\nu=1$}. Assuming that the $E_{Z}$ in our sample  is
enhanced by the same factor, we obtain $g\approx 0.04$, which is
the same value as reported by Shukla \textit{et
al.}\cite{Shukla00}. In other words, it appears  that the nature of
 the interactions responsible for the gap enhancement at $\nu =1$ is 
similar in both GaAs and \GAl QWs.

Next, we will discuss the origin of the linewidth broadening at
low temperatures. Two possible mechanisms  are
sample inhomogeneities and/or large skyrmions. Dynamic
measurements of the spin-spin decoherence time ($T_{2}^{-1}$) are
employed to differentiate between these. We have observed a very fast decay rate 
$(T_{2})^{-1}$ of the amplitude of the spin echo  (shorter than the dead time for  signal
detection, $\sim 10\, \mu $s), preventing us from  pursuing  more
quantitative studies. Such dynamic fluctuations cannot be induced
by the static sample inhomogeneities, 
 since we were able to record spectra using free induction decay. 
 Therefore, it is likely that
our observations reveal some aspect of skyrmion physics. Mainly, a
very fast $(T_{2})^{-1}$  rate reveals that the spin polarization
distribution presents strong dynamic fluctuations at the time scale of the spacing 
between the RF pulses (a few microseconds).  These dynamic
fluctuations can be associated with spin-textured domains that are
  dynamic themselves and/or whose $\nu$ distribution moves
rapidly on the NMR time scale. We measured comparable
$\mathcal{P}$ and full width at half maximum (FWHM) values at $\nu =1$ and $\nu =1\pm 0.1$, which favors
 the dynamic $\nu$ distribution scenario. Furthermore, in such a scenario, 
$\mathcal{P}$ would   be significantly diminished as is observed here and
  discussed in \mbox{Ref. \cite{Melinte}}. The most likely
scenario is that dynamic variations of $\nu $\ are realized
through skyrmion motion, meaning that they are concomitant with
dynamic skyrmions.

This dynamic picture is endorsed by the temperature dependence of
the linewidth (FWHM). Well-known NMR phenomena of motional
narrowing implying that the FWHM is strongly affected by the
dynamics of the nuclei\cite{AbragamBook}. Since the nuclei are
spatially localized in the lattice, any observed dynamical effect
must be related to the dynamics of electrons in QWs, i.e., to the
motion of delocalized quasiparticles or skyrmions. The FWHM depicts
an evolution of the skyrmion dynamics from the motionally-narrowed
to the frozen regimes with decreasing temperature
\cite{Khandelwal01}. The slow skyrmion dynamics  tend  to
increase the linewidth. However, the FWHM should ultimately
decrease at very low $T$ since the exponentially small number of
skyrmions cancels out the effect of the slow dynamics
\cite{Villares04}. Therefore, the temperature dependence of the FWHM
is expected to be nonmonotonic. As portrayed in \mbox{Fig.  \ref{GSubLW}},
the FWHM continuously increases with decreasing temperature
implying a progressive slowing down of skyrmion dynamics. Lack of the
peak in the FWHM \textit{versus} $T$ graph indicates that skyrmions do
not become fully localized \cite{Villares04} down to $T=65$ mK, the
lowest $T$ in our experiment. We remark further that the lack of the peak
 argues against the impurity scenario, since one expects impurities to
be effective in pinning quasiparticles at such a low temperature.


\section{Conclusions}

In a quantum wells  sample
with a small $g$ tensor  we have employed a special NMR technique to allow the distinction
between a weak QW signal and the barrier signal. 
We determined the polarization of the
electrons in the QWs and its temperature evolution. The measurements
directly confirm the theoretical prediction (\mbox{$g \approx +
0.04$}) of a  small, positive value for  the $g$ 
factor \cite{Shukla00}. A very broad distribution of spin polarization  was found
in the sample. This can be attributed to sample inhomogeneities
 or to large skyrmions. The latter possibility is favored by
observation of a very fast $T_2^{-1}$ rate, indicating that the
spin polarization distribution is dynamic   rather  than static.
\newline

We are grateful to Y. Tokunaga  for useful discussions. Special
thanks to P. van de Linden for technical assistance. 
This work was partially  supported
by 
the Grenoble High Magnetic Field Laboratory, under European 
Community contract RITA-CT-2003-505474, and  by the NSF.

\vspace{0.5cm} 

\end{document}